\documentclass[twocolumn,preprintnumbers,amsmath,amssymb]{revtex4}
\usepackage{graphicx}
\usepackage{dcolumn}
\usepackage{bm}
\usepackage{hyperref} 
\usepackage{multirow} 
\usepackage{soul}
\newcommand\beq{\begin{equation}}
\newcommand\eeq{\end{equation}}
\newcommand\bea{\begin{eqnarray}}
\newcommand\eea{\end{eqnarray}}
\usepackage{wrapfig}
\newcommand{\nonum}{\nonumber}

\begin{document}

\title{\bf  Magnetic susceptibility of alkali-TCNQ salts and extended Hubbard models with  bond order and charge density wave phases \\}
\author{\bf  Manoranjan Kumar, Benjamin J. Topham,   Rui Hui Yu, 
  Quoc Binh Dang Ha\footnote{Present address: University College Dublin,  Dublin Ireland 01 716 7777},  Zolt\'an G. Soos\footnote{Electronic mail:soos@princeton.edu}}
\address{  Department of Chemistry, Princeton University, Princeton NJ 08544 \\}
\date{\today}

\begin{abstract}
The molar spin susceptibilities $\chi(T)$ of Na-TCNQ, K-TCNQ and Rb-TCNQ(II) 
are fit quantitatively to 450 K in terms of half-filled bands of three 
one-dimensional Hubbard models with extended interactions using exact results 
for finite systems. All three models have bond order wave (BOW) and charge density 
wave (CDW) phases with boundary $V = V_c(U)$ for nearest-neighbor interaction $V$ 
and on-site repulsion $U$. At high $T$, all three salts have regular stacks of $\rm TCNQ^-$ 
anion radicals. The $\chi(T)$ fits place Na and K in the CDW phase and Rb(II) in 
the BOW phase with $V \approx V_c$. The Na and K salts have dimerized stacks at $T < T_d$ 
while Rb(II) has regular stacks at 100K. The $\chi(T)$ analysis extends to dimerized 
stacks and to dimerization fluctuations in Rb(II). The three models yield consistent 
values of $U$, $V$ and transfer integrals $t$ for closely related $\rm TCNQ^-$ stacks. 
Model parameters based on $\chi(T)$ are smaller than those from optical data that in 
turn are considerably reduced by electronic polarization from quantum chemical 
calculation of $U$, $V$ and $t$ on adjacent $\rm TCNQ^-$ ions. The $\chi(T)$ 
analysis shows that fully relaxed states have reduced model parameters compared to 
optical or vibration spectra of dimerized or regular $\rm TCNQ^-$ stacks.
 
\vskip .4 true cm
\end{abstract}

\maketitle

\section{Introduction}
The strong $\pi$-acceptor A = TCNQ (tetracyano-quinodimethane) forms 
an extensive series of ion-radical salts \cite{r1,r2,r3} with 
closed-shell inorganic ions as well as charge-transfer (CT) 
complexes with $\pi$-donors such as D = TTF (tetrathiafulvalene). 
The high conductivity and phase transitions of TTF-TCNQ on cooling 
were thoroughly investigated as an important step towards the realization 
of organic superconductivity \cite{r4}. TCNQ salts crystallize in face-to-face 
stacks that immediately rationalize their quasi-one-dimensional (1D) 
electronic structure. Endres has reviewed the many structural motifs of 
1D stacks \cite{r5}. We consider in this paper the magnetic properties of 
``simple'' 1:1 alkali-TCNQ salts with half-filled stacks of $\rm A^-$ radical ions. 
Complex salts with stoichiometry such as 1:2 or 2:3 have less than half-filled stacks; 
they are semiconductors with higher conductivity than simple salts. Hubbard and related 
models are the standard approach to TCNQ salts or CT complexes \cite{r1,r2,r3,r4}. 
Each molecule in a stack is a site with a single frontier orbital, the 
lowest unoccupied orbital of A or the highest occupied orbital of D.\\ 

Heisenberg spin chains were initially applied to the magnetic properties of 
TCNQ salts \cite{r6}, especially to dimerized stacks whose elementary excitations 
are triplet spin excitons. Subsequently, Hubbard models \cite{r1,r2,r3,r4} opened the 
way to discuss optical and electrical as well as magnetic properties. Limited understanding 
of 1D models hampered early treatments. Theoretical and numerical advances now make 
it possible to treat the spin susceptibility of 1D models almost quantitatively. 
Alkali-TCNQ salts offer the possibility of joint modeling of magnetic, optical and 
vibrational properties. An interesting consequence reported below is that different 
model parameters are needed for magnetic and optical properties.\\
 
There is considerable literature on K-TCNQ or Na-TCNQ, recently in connection 
with photo-induced phase transitions \cite{r7,r8}. They form \cite{r5} regular stacks with $\rm A^-$ 
at inversion centers at high $T$, dimerized stacks with two $\rm A^-$ per repeat unit at low $T$. 
The transitions are \cite{r9} at $T_d$ = 348 K and 395 K, respectively, for $\rm Na$ and 
K-TCNQ. Torrance \cite{r10} and others \cite{r11} sought to model $T_d$ as a spin-Peierls 
transition, as discussed in the review of Bray et al. \cite{r12} who noted that 
such high $T_d$ requires unacceptably large exchange constants. 
The transitions have some 3D character since the cations also dimerize \cite{r13,r14}. 
We model the molar spin susceptibility $\chi (T)$ of the Na and K salts at $T > T_d$ 
using regular stacks. We also consider $\chi (T)$ of dimerized stacks for $T < T_d$ 
without, however, treating the transition. There are two Rb salts: Rb-TCNQ(I) is 
strongly dimerized \cite{r15} at 300 K while Rb-TCNQ(II) has regular stacks \cite{r16,r17} 
with $A^-$ at inversion centers at both 100 and 295 K. The recent 100 K structure \cite{r17} 
rules out a dimerization transition around 220 K that was inferred from 
magnetic susceptibility \cite{r18} and infrared \cite{r19} data. We 
reinterpret these observations. Regular stacks make Rb-TCNQ(II) the best target for modeling $\chi (T)$. \\

\begin{figure}[h]
\begin {center}
\hspace*{-0cm}{\includegraphics[width=8.5cm,height=10.0cm,angle=90]{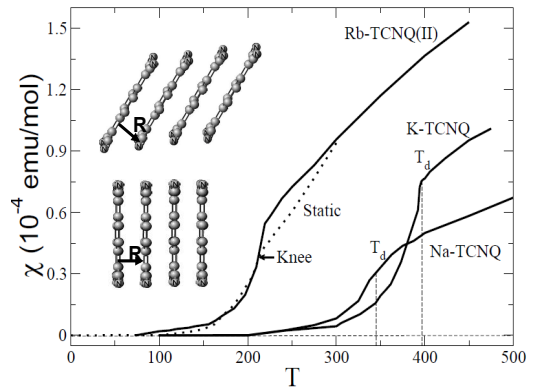}} \\
\caption{Solid lines: molar magnetic susceptibility  of alkali-TCNQ salts 
based on electron spin resonance intensity from ref. \cite{r18}; dotted line, 
static susceptibility from ref. \cite{r17}. The Na and K salts dimerize at $T_d$ 
and have regular face-to-face stacks for $T > T_d$ with separation 
R between molecular planes. The Rb salt has regular ring-over-external-bond 
stacks down to 100 K and a knee around 220 K. }
\label{fig1}
\end {center}
\end{figure}

Fig. \ref{fig1} shows the molar spin susceptibilities of $ \rm Na$, $\rm K$ and Rb-TCNQ(II). 
The K and Na data are integrated electron spin resonance (esr)  
of Vegter and Kommandeur \cite{r18}, who identified the transitions. Dimerization opens 
a magnetic gap $E_m > 0$ and rationalizes reduced $\chi (T) $ that vanishes at $T = 0$, 
whether or not $\chi (T)$ can be modeled. Crystal data \cite{r13,r14} at $T > T_d$ 
indicate eclipsed (ring over ring, Fig. \ref{fig1}) stacks with $\rm TCNQ^-$ at 
inversion centers and interplanar separation R(Na) = 3.385 $\AA$ at 353 K, R(K) = 3.479 $\AA$ 
at 413 K. The solid line for Rb(II) is esr intensity \cite{r18}. The dotted line is static 
susceptibility \cite{r17} corrected for diamagnetism. The measurements agree at 
300 K and both have a knee around 220 K, less prominently in static susceptibility. 
The structure has slipped stacks \cite{r17} (ring over external bond, Fig. \ref{fig1}) 
of $ \rm TCNQ^-$ at inversion centers with R = 3.174 and 3.241 $\AA$ at 100 and 295 K, 
respectively. These regular stacks clearly have large $E_m$. They are not compatible 
with finite $\chi(0)$ at $T = 0$ and $E_m = 0$ in regular Heisenberg \cite{r20} or Hubbard \cite{r21} chains.\\
 
The 1D extended Hubbard model \cite{r22} (EHM, Eq. \ref{eq1} below) has nearest-neighbor 
interaction $V$ in addition to on-site repulsion $U > 0$. Increasing $V$ in a 
half-filled EHM induces a transition to a charge density wave (CDW) phase \cite{r22}. 
The CDW boundary is at $V_c(t = 0) = U/\alpha_M$ in the atomic limit of $U >> t$, 
where $t$ is electron transfer between adjacent sites and $\alpha_M = 2$ 
for the EHM is the Madelung constant of the lattice. The CDW transition 
is closely related to the neutral-ionic transition of CT salts from largely 
neutral DADA stacks to largely ionic $\rm D^+A^-D^+A^-$ stacks \cite{r23,r24,r25}. 
In either case, the ground state (gs) undergoes a first-order quantum transition at 
small $t/U$ or a continuous transition when $t/U$ exceeds a critical value, 
or when $U < U^*$. Nakamura \cite{r26} recognized that the EHM with $U < U^*$ 
has a narrow bond order wave (BOW) phase between the CDW phase at $V_c(U) > U/2$ 
and the spin-fluid phase with $E_m = 0$ at $V_s(U) < U/2$. The BOW phase has finite $E_m$ 
in a regular stack and broken inversion symmetry $\rm C_i$ at sites. 
Subsequent studies \cite{r27,r28,r29} confirmed the BOW phase of the 
EHM and sought accurate values of $V_s(U)$, $V_c(U)$ and $U^*$. 
We recently characterized the BOW phase of the EHM and related broken $C_i$ 
symmetry to electronic solitons \cite{r30}. Finite $E_m$ in regular stacks 
is an attractive way to rationalize $\chi(T)$ in Fig. \ref{fig1}, and we have 
proposed that Rb-TCNQ(II) is a BOW phase system \cite{r17,r31}. 

In this paper, we model the spin susceptibility of Na, K and Rb-TCNQ(II) 
quantitatively with the EHM and related models with more realistic Coulomb interactions. 
We find the Na and K salts to be in the CDW phase with $V$ slightly greater than $V_c(U)$ and 
the Rb(II) salt to be just on the BOW side of $V_c(U)$. Modeling $\chi(T)$ is both challenging 
and decisive for several reasons. First, the full electronic spectrum is required, not just 
the ground state. Second, comparison with experiment is absolute since the magnitude of $\chi(T)$ 
follows without scaling in $\pi$-radicals with weak spin-orbit coupling. Third, all 
three salts have 1D stacks of $\rm A^- = TCNQ^-$ with similar $U$ and other parameters on 
physical grounds. To the best of our knowledge, Hubbard models have not been applied 
quantitatively to both magnetic and optical/vibronic properties of the same system. 
1:1 alkali-TCNQ salts provide such an opportunity.\\

The paper is organized as follows. We present in Section II 
the spin susceptibility of Hubbard-type models near the boundary $V_c(U)$ 
between the BOW and CDW phases. The magnetic gap $E_m(V)$ to the lowest 
triplet state increases rapidly at $V \approx V_c$. 
In Section III we model the $\chi(T)$ data in Fig. \ref{fig1} with similar parameters for $\rm TCNQ^-$ 
stacks in related but not identical crystals. We compute model parameters in Section IV 
for individual $\rm TCNQ^-$ or for adjacent $\rm TCNQ^-$. These parameters are 
reduced substantially in crystals, more so for magnetic than for optical or 
vibrational properties. The Discussion briefly addresses the parameters of H\"{u}ckel, 
Hubbard or other semiempirical models.\\ 
\section{BOW/CDW boundary}
We consider a half-filled extended Hubbard model \cite{r22} (EHM) in 
1D and extend it to second-neighbor interactions $V_2 = \gamma V$. The EEHM with $\gamma  > 0$ is
\begin{eqnarray}
H(\gamma) & =& \sum^N_{p=1,\sigma} -t(  a^{\dagger}_{p,\sigma}  a_{p+1,\sigma} + h.c) \nonum \\
       & + & \sum^N_{p=1}  U n_p(  n_p-1)/2 \nonum \\
&+& \sum^N_{p=1} V n_p (n_{p+1}+ \gamma n_{p+2} \label{eq1})
\end{eqnarray}
\noindent The first term describes electron transfer between adjacent sites with retention of spin $\sigma$. 
Regular stacks in this Section have equal $t$’s taken as $t = 1$. The number operator is $n_p$. 
The last two terms are on-site repulsion $U > 0$, nearest-neighbor interaction $V$ and 
second-neighbor interaction $\gamma V$. The spin fluid phase with $n_p = 1$ 
at all sites is the gs for small $V$ while the charge density wave (CDW) with 
two electrons per site on one sublattice is the gs for large $V$. The CDW boundary is $V_c(t = 0) = U/\alpha_M(\gamma)$, 
where $\alpha_M(\gamma) = 2(1 - \gamma)$ is the 1D Madelung constant. 
As recognized from the beginning \cite{r32,r6}, electrostatic interactions are 3D and 
ion-radical organic salts have $\alpha_M \approx 1.5$. Point charges in 1D lead to $\rm \alpha_M = 2ln2$. Physical considerations set $\alpha_M=1.5$ rather than $\alpha_M(EHM) = 2$. \\

Finite $t$ in a regular stack leads to a narrow BOW phase \cite{r26} between $V_s < U/\alpha_M$ 
and $V_c > U/\alpha_M$ for $U < U^*$, with\cite{r29} $U^* \approx 7t$ for the EHM. Smaller $\alpha_M$ 
gives a less cooperative CDW transition and extends the BOW phase to higher $U^*$. 
The point charge model (PCM) with long-range Coulomb interactions 
$V_n = V/n$ in Eq. \ref{eq1} has \cite{r33} $U^*(\rm PCM)$$~ \approx ~ 10t$. 
By the same analysis, we estimate that the EEHM with $\gamma = 0.2$ in Eq. \ref{eq1} 
has $U^* \approx 9t$. Quantum chemical evaluation \cite{r34} of $U$ and $V$ 
places alkali-TCNQ salts at the CDW boundary and imposes the constraint $V \approx V_c$ in Eq. \ref{eq1}. \\

The symmetry properties of $H(\gamma)$ are the same for spin-independent interactions. 
Total spin $S$ is conserved and $E_m$ is from the singlet gs to the lowest triplet state. 
The half-filled band has electron-hole symmetry $J = \pm 1$. We define $E_J$ as the excitation 
energy to the lowest singlet with opposite $J$ from the gs. A regular stack has inversion 
symmetry $C_i$ at sites that we label as $\sigma = \pm 1$ and define $E_{\sigma}$ 
as excitation to lowest singlet with opposite $\sigma$ from the gs. 
The energy thresholds $E_m$, $E_J$ and $E_{\sigma}$ of extended stacks are not known exactly for 
$V > 0$ in Eq. \ref{eq1}. \\

We consider $V \approx V_c(N)$ and use valence bond methods \cite{r35} to solve $H(\gamma)$ 
exactly for $N = 4n$ or $4n + 2$ sites with periodic or antiperiodic boundary conditions, 
respectively. Low-energy excitations are accessible up to $N = 16$, and the full spectrum to $N = 10$. 
At constant $U$ and $t$, the condition $E_m(V) = E_{\sigma}(V)$ gives $V_s(N)$ while $E_{\sigma}(V) = E_J(V)$ 
gives $V_c(N)$. We also define $V_1(N)$ where $E_{\sigma}(V) = 0$ and the 
degenerate gs in the BOW phase can be explicitly constructed as linear combinations of $\sigma = \pm 1$ 
functions \cite{r30}. Table \ref{tb1} lists $V$’s in units of $t$ for $\gamma = 0.2$ ($\alpha_M = 1.6$) 
and $U = 8t$ in Eq. \ref{eq1} up to $N = 16$. The $V$’s cluster as expected about $U/\alpha_M = 5.0$. 
Their weak $N$ dependence makes it possible to extrapolate to the 
extended system as discussed \cite{r36,r37} in connection with a frustrated spin chain. 
 We have computed $V_s(N)$, $V_1(N)$ and $V_c(N)$ of all three models (EHM, EEHM, PCM) as functions of $U < U^*$ 
and have previously reported \cite{r30} EHM values at $U = 4t$.\\

\begin{table}
\caption { Boundaries $V_s$ and $V_c$ of the BOW phase of $H(\gamma)$, Eq. \ref{eq1},
with $N$ sites, $t = 1$, $U = 8$, $\gamma = 0.2$, and periodic (antiperiodic) boundary
conditions for $N = 4n (4n + 2)$ based on excitation thresholds $E_m$, $E_{\sigma}$, $E_J$.}
\begin{tabular}{c|c|c|c} \hline
~~$N$~~ & ~~~$V_s (E_{\sigma} = E_m)$~~~ & ~~~$V_1(E_{\sigma} = 0)$~~~& ~~~$ V_c (E_{\sigma} = E_J)$~  \\\hline
8 &  4.834 & 5.105 & 5.199  \\
10 & 4.908 & 5.139 & 5.201 \\
12 & 4.908 & 5.139 &5.201 \\
14 & 4.932  &5.149  & 5.202\\
16 & 4.952 &5.157 & 5.203  \\\hline

\end{tabular}
\label{tb1}
\end{table}
 
The magnetic gap $E_m$ dominates $\chi(T)$ as $T \rightarrow 0$. 
It opens \cite{r26,r33} slowly at $V_s$, remains small at $V_1$ 
and grows rapidly on crossing the CDW boundary at $V_c$. The size dependence of $E_m$ 
in Table \ref{tb2} is for the EEHM at $U = 8t$. Decreasing $E_m(N)$ is found 
in spin or Hubbard chains with $E_m = 0$ in the extended system. 
Instead, $E_m$ increases with $N$ in all three models when $V$ slightly exceeds $V_c(N)$. 
A density matrix renormalization group (DMRG) calculation \cite{r33} 
for the EHM at $U = 4t$ shows increasing $E_m(N,V)$ for $N > 30$ at $V = V_c(N)$. 
Hence $E_m$ of the extended system may exceed the $N = 8$ gaps that we use below for $V \approx V_c$ 
in the BOW phase for $V > V_c$ in the CDW phase.\\

\begin{table}
\begin{center}
\caption {Magnetic gap $E_m(V)$ to the lowest triplet of $H(\gamma)$, Eq. \ref{eq1}, 
with $N$ sites, $t = 1$, $U = 8$, $\gamma = 0.2$, and periodic (antiperiodic) 
boundary conditions for $N = 4n$ $(4n + 2)$.}
\begin{tabular}{c|c|c|c} \hline
$N$&~~~~~  $E_m(V_1)$~~ ~~~ &~~~~~ $E_m(V_c)$~~~~~ & ~~~~$E_m(V_c+0.3)$ \\\hline
8  &0.325  &0.496  & 1.247\\
10 &0.268  &0.451  & 1.293\\
12 &0.263  &0.413  & 1.340\\
14 &0.257  &0.384  & 1.378\\
16 &0.242  &0.364  & 1.403  \\\hline
\end{tabular}
\end{center}
\label{tb2}
\end{table}
\begin{figure}[h]
\begin {center}
\hspace*{-0cm}{\includegraphics[width=7.0cm,height=9.5cm,angle=-90]{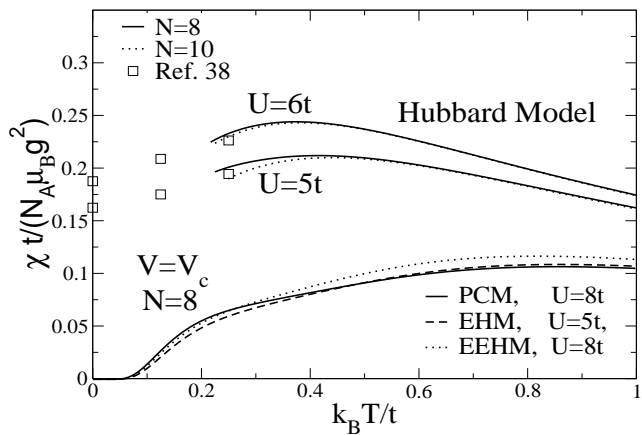}} \\
\caption{Molar magnetic susceptibility $\chi$ of Hubbard-type models, Eq. \ref{eq1}, with finite $N$. 
The $U = 5t$ and $6t$, $V = 0$ curves for $N = 8$ and 10 match
the extended results (open symbols) from ref. \cite{r38} for $k_BT > 0.3t$.
The other curves have $V = V_c$. The EHM has $U = 5t$, the EEHM has $U = 8t$ and $\chi = 0.2$
in Eq. \ref{eq1}, and the PCM has $U = 8t$ and Coulomb interactions $V_n = V/n$. }
\label{fig2}
\end {center}
\end{figure}

We compute the full spectrum of $H(\gamma)$ for $N = 8$ (10) sites with 
periodic (antiperiodic) boundary conditions \cite{r30}. 
Standard methods give the partition function $Q$ and molar spin susceptibility 
$\chi(T)$. Fig. \ref{fig2} shows $\chi(x)$ as a function of reduced temperature 
$x = k_BT/t$, where $k_B$ is the Boltzmann constant.
 Since $g( \rm TCNQ^-) \approx  2.00236$, the free-electron value, $\chi(x)$ is directly related to 
Avogadro's number $N_A$ and the Bohr magneton $\mu_B$. 
J\"{u}ttner et al. \cite{r38} obtained quantitative $\chi(x)$ 
for the Hubbard model with $V = 0$ in Eq. \ref{eq1} and finite $\chi(0)$; 
their results for $U = 5t$ and $6t$ are shown by open symbols in Fig. \ref{fig2}. 
The lines are exact $N = 8$ and 10 results that for $x > 0.3$ coincide with the 
extended chain within our ability to read graphs. The other $\chi(x)$ 
curves in Fig. \ref{fig2} are for $N = 8$ with periodic boundary conditions and $V = V_c$. 
We find similar $\chi(x)$ for the EHM with $U = 5t$, 
the EEHM with $U = 8t$, $\gamma = 0.2$ and the PCM with $U = 8t$. Small $V$ is conveniently 
approximated as a Hubbard model with an effective $U_e = U - V$. 
This rationalizes reduced $\chi(x)$ with increasing $V$, but not the qualitative change of 
$\chi(0) = 0$ due to finite $E_m$ in the BOW or CDW phase. Large $t \approx 1000 K$ and 
limited thermal stability of ion-radical oganic solids limits $\chi(x)$ to $x < 0.5$.

\section{ Magnetic susceptibility}

In this Section, we model $\chi(T)$ data in Fig. \ref{fig1} using regular stacks for Rb(II) 
and for K and Na at $T > T_d$. The first terms of Eq. \ref{eq1} for dimerized stacks at $T < T_d$ has transfer integrals

\begin{eqnarray}
t_p &=& -(1+(-1)^p \delta) 
\label{eq2}
\end{eqnarray}
\begin{figure}[h]
\begin {center}
\hspace*{-0cm}{\includegraphics[width=7.0cm,height=9.5cm,angle=-90]{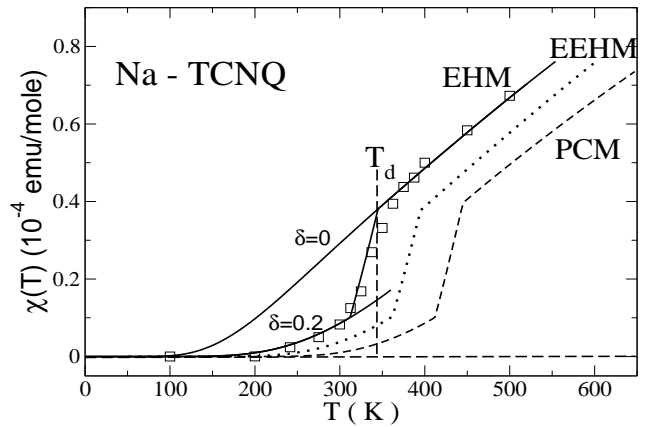}} \\
\caption{  Three $\chi_M(T)$ fits shifted by 50 K for clarity of the Na-TCNQ data (open symbols)
in Fig. \ref{fig1} for Hubbard-type models in Eq. \ref{eq1} with $N = 8$ and
parameters in Table \ref{tb3}.
The stacks are regular ($\delta = 0$) for $T > T_d$, dimerized ($\delta = 0.20$) at low $T$
and interpolated using Eq. \ref{eq3} in between.}
\label{fig3}
\end {center}
\end{figure}
along the stack. We did not change $V$ in dimerized stacks. Since all three salts have 
$\rm TCNQ^-$ stacks, similar $U$ is expected on physical grounds, and we have sought 
similar $U < U^*$ without strictly enforcing the constraint. It soon became apparent 
that $\chi(T)$ for $T > T_d$ requires $V > V_c$. We studied the 
PCM with $V_n = V/n$ and EEHM with $\gamma = 0.2$ in Eq. \ref{eq1} in addition 
to the EHM ($\gamma = 0$) in part to search for a fit with $V \leq V_c$ 
and in part to probe the dependence of $t$ and $U$ on the model. The following $\chi(T)$ 
calculations are all for $N = 8$ sites with periodic boundary conditions in Eq. \ref{eq1}. 
The experimental data in Fig. \ref{fig1} are now shown as open symbols.\\

We start with $\chi(T)$ for Na-TCNQ in Fig. \ref{fig3} and obtain good fits for $T > T_d$ 
for the EHM with $U/t = 4$, $t = 0.097$ eV and $V/t = V_c + 0.19$. The $\chi(T)$ 
results for the EEHM and PCM are displaced by 50 and 100 K, respectively, for clarity. 
They are equally good for the $t$, $U$ and $V$ parameters listed in Table \ref{tb3}. 
All three models return $t \approx 0.10$ eV and $V$ slightly larger than $V_c$. Good $\chi(T)$ 
fits to $T = 310$ K in the dimerized phase are shown in Fig. \ref{fig3} with $\delta = 0.20$ 
in Eq. \ref{eq2} and the same $t$, $U$ and V. Konno and Saito \cite{r13} followed the 
temperature dependence of the Na-TCNQ crystal structure and found a coexistence region. 
The regular phase for $T > T_d$ = 345 K appears already at $T = 332$ K and 
grows at the expense of the dimerized phase that disappears at $T_d$. The $\chi(T)$ 
fits in Fig. \ref{fig3} between $T_1 = 310$ K and $T_d$ are linear interpolations according to

\begin{eqnarray}
\chi(T)=\frac{T-T_1}{T_d-T_1}\chi(T,\delta=0)+\frac{T_d-T}{T_d-T_1}\chi(T,\delta=0.2)
\label{eq3}
\end{eqnarray}
The coexistence region is 10 K wider in the fit. Terauchi \cite{r9} studied the intensity 
of selected superlattice reflections for $T < T_d$ in both Na and 
K-TCNQ. The intensity is proportional to ${\delta(T)}^2$ and decreases 
linearly as $T_d- T$ near $T_d$. The susceptibility between 320-345 K can 
also be modeled as variable $\delta(T)$. \\ 
 
\begin{figure}[h]
\begin {center}
\hspace*{-0cm}{\includegraphics[width=7.0cm,height=9.5cm,angle=-90]{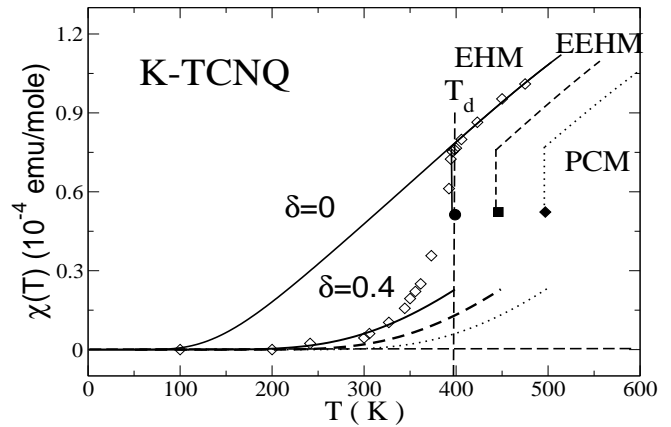}} \\
\caption{ Three $\chi_M(T)$ fits shifted by 50 K for clarity of the 
K-TCNQ data (open symbols) in Fig. \ref{fig1} for Hubbard-type models in 
Eq. \ref{eq1} with $N = 8$ and parameters in Table \ref{tb3}. 
The stacks are regular ($\delta = 0$) for $T > T_d$, dimerized ($\delta = 0.40$) 
at low $T$ and jump to $\delta/2$ at $T_d$.}
\label{fig4}
\end {center}
\end{figure}

Fig. \ref{fig4} shows $\chi(T)$ fits for K-TCNQ, again displaced by 50 K for clarity and
again in the CDW phase with $V > V_c$ for $T > T_d$ = 398 K. The K-TCNQ parameters $t$, $U$ and $V$
are in Table \ref{tb3}. The same parameters and $\delta = 0.40$ agree with experiment up to 350 K.
There is no evidence of coexisting phases. The intensity of superlattice reflections decreases over an 
80 K interval and changes discontinuously from $\delta/2$ to 0 at the transition \cite{r9}. The solid points 
are calculated $\chi(T_d, \delta/2)$. Agreement at $\chi(T_d)$ indicates that $\chi(T)$ between 
350 K and $T_d$ can be fit with variable $\delta(T)$ in these models. Smaller $t(K) \approx 0.08 eV$ is 
consistent with larger R in K-TCNQ. \\

\begin{table}
\caption {Parameters for the spin susceptibility of Na, K and Rb-TCNQ(II) in Figs. \ref{fig3},\ref{fig4},\ref{fig5}}
\begin{center}
\begin{tabular}{c|c|c|c|c} \hline
Salt &~~~~Model~~~~ &~~~~ $t(eV)$~~~~ &~~~~ $U(eV)$ ~~~~& ~~~~$V(eV)$\\
  &        &          &         &       \\\hline 
  & EHM    &  0.0956  &  0.383  &  0.214\\
Na& EEHM   &  0.0969  &  0.630  &  0.434\\
  & PCM    &  0.0965  &  0.627  &  0.492\\\hline
   &EHM    &  0.0780  &  0.370  &  0.211\\
K  &EEHM   &  0.0801  &  0.601  &  0.406\\
   &PCM    &  0.0758  &  0.569  &  0.440\\\hline
   &EHM    &  0.0745  &  0.373  & 0.199\\
 Rb &EEHM  &  0.0767  &  0.614  &  0.399\\
   &PCM    &  0.0707  &  0.601  &  0.440\\\hline
\end{tabular}
\end{center}
\label{tb3}
\end{table}

Figure 5 shows $\chi(T)$ fits for Rb-TCNQ(II) for the three models displaced by 50 K. 
We took $V = V_c(8)$ at the upper limit of the BOW phase and set $E_{\sigma} = 0$. 
The $\delta = 0$ fit for regular stacks is markedly improved by slightly increasing $E_m$ 
beyond $E_m(8)/t$ in Table \ref{tb2}, by $0.07t$ for the EHM and by $0.10t$ for PCM and EEHM. 
Finite-size effects are critical in view of other evidence \cite{r17,r30} for broken $ \rm C_i$ 
symmetry in Rb-TCNQ(II), which implies $V \leq V_c$. By contrast, finite-size effects for the 
Na or K salts are absorbed in $V > V_c$. Good $\delta = 0$ fits are obtained down to $T \approx 250$ K 
with the $t$ and $U$ parameters in Table \ref{tb3}. 
The esr intensity in Fig. \ref{fig1} has a pronounced knee around $T_{kn} \approx 220$ K. 
The knee is less prominent in the static susceptibility. Dimerization is ruled out by the 100 K structure, 
which has regular stacks and the 300 K space group \cite{r17}. \\

An adiabatic (Born-Oppenheimer) approximation for the lattice is typically invoked 
to model the Peierls \cite{r39} or spin-Peierls \cite{r12,r40} instability of 1D 
systems, although quantum fluctuations\cite{r41} are important for small $\delta(0)$ at 
$T$ = 0. The BOW phase has finite $\delta(0)$ for linear electron-phonon (e-ph) coupling 
$\alpha$ to a harmonic lattice \cite{r30}, where $\alpha= (dt/du)_0$ is the 
first term of the Taylor expansion of $t(R+u)$. 
The electronic gs energy per site in units of $t$ has a cusp \cite{r30}
\begin{eqnarray}
\epsilon_0(\delta)-e_0(0)=-B(V)|\delta|+O(\delta^2)
\label{eq4}
\end{eqnarray}
\begin{figure}[h]
\begin {center}
\hspace*{-0cm}{\includegraphics[width=7.0cm,height=9.5cm,angle=-90]{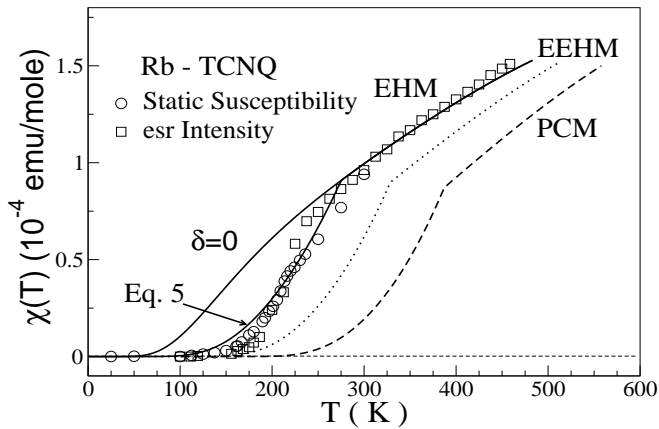}} \\
\caption{Three $\chi(T)$ fits shifted by 50 K for clarity of the Rb-TCNQ(II) data (open symbols) in
Fig. \ref{fig1} for Hubbard-type models in Eq. \ref{eq1} with $N = 8$ and parameters in Table \ref{tb3}.
The stacks are regular ($\delta = 0$). Spin solitons with width $2\xi = 60$ in Eq. \ref{eq5} are used at low $T$.} 
\label{fig5}
\end {center}
\end{figure}

\noindent where $B(V)$ is the order parameter of the BOW phase. $B(V)\approx 0.4$ is the bond-order 
difference at $V = V_1(U)$ in Table \ref{tb1} for all three models for the $U$s in 
Table \ref{tb3}. For comparison, a half-filled band of free electrons with $\delta = \pm 0.1$ 
has comparable $B(\delta)$ for partial double and single bonds.\\ 

The BOW phase has long-range order that cannot persist for $T > 0$ in 1D systems. As discussed by Su, 
Schrieffer and Heeger \cite{r39} for free electrons, the extended system at low $T$ has regions with 
reversed $\delta(0)$ that are separated by topological solitons whose width $2\xi$ goes as $1/\delta(0)$. 
Spin solitons are also found numerically in the BOW phase \cite{r30} of the EHM or 
in the magnetic properties of organic ion-radical salts \cite{r42}.\\

We consider dimerization fluctuations in the BOW phase. This regime has equal densities $\rho(T)$ of spin solitons 
and dimerized segments with successively $\pm \delta(0)$ 
in Eq. \ref{eq4}. We approximate each $S = 1/2$ soliton by 
 a regular region of $2\xi$ sites in an otherwise dimerized stack. 
Since $E_m$ is not degenerate, $E_m(V,\delta(0))/t$ initially increases as $B(V)|\delta(0)| N$ 
due to the cusp in Eq. \ref{eq4}, as found directly \cite{r30} up to $N = 16$ at $V = V_1(N)$ 
where $E_{\sigma} = 0$. Such size dependence cannot go on indefinitely. 
It suffices for our purposes to note that $\delta(0) < 0.10$ generates large $E_m$ 
with negligible $\chi(T)$ at low $T$ in dimerized regions between solitons. 
Such regions decrease with increasing $T$ and  vanish at  $2\xi \rho(T') = 1$ 
when the stack is regular everywhere.\\ 

In this approximation, dimerization fluctuations reduce $\chi(T)$ for $T < T'$. 
The soliton density $\rho(T) = \chi(T)T/C$ follows directly from the molar Curie constant $C = N_Ag^2\mu^2_B$ 
of noninteracting spins. The knee region in Fig. \ref{fig5} up to $T’$ is modeled as
\begin{eqnarray}
\chi(T < T', \xi ) &=& 2\xi \rho(T) \chi(T) \label{eq5}
\end{eqnarray}
\noindent with $2\xi = 60$. The fit is adequate for the simple treatment of fluctuations. The choice of $2\xi = 60$ 
gives $T' \approx 250$ K, somewhat higher than experiment. The same soliton width accounts for the $T$ 
dependence of the infrared intensity of a totally symmetric $\rm TCNQ^-$ vibration \cite{r30}. 
Such IR data is decisive evidence \cite{r43,r44} for broken $\rm C_i$ symmetry, whether due to $B(\delta)$ 
in well-characterized K-TCNQ stacks \cite{r45} at 300 K or to finite $B(V)$ in a BOW phase.\\ 

It remains to reconcile dimerization fluctuations at low $T$ with the X-ray data for 
regular stacks at 100 K and thermal ellipsoids that conservatively limit \cite{r46}  $ \rm R_+ - R_-$ = $2u < 0.04 \AA$. 
To be detectable, $u$ must exceed zero-point motions. The stack at 100 K has small $\rho(T)$ 
that prevents long range order. Soliton motion modulates R as $\delta(0) = \alpha u/t$, 
and the magnitude of $\alpha/t$ determines whether $\delta(0)$ is consistent with X-ray data. \\ 

We conclude this Section by assessing the parameters in Table \ref{tb3}. 
Three models (EHM, EEHM, PCM) with narrow BOW phases have been applied to three $\rm TCNQ$ 
salts (Na, K, Rb(II)). The CT integral $t$ of regular stacks depends on 
overlap, as sketched in Fig. \ref{fig1}, and on separation between $\rm TCNQ^-$ planes. 
It is reassuring that the models return identical $t$ to better than 10 \% 
with $t(Na) > t(K) > t(Rb)$ in an unconstrained fit. We sought similar $U$ 
in $\rm TCNQ^-$ stacks. The $U$’s in Table \ref{tb3} are identical within 5\% for each model. 
The BOW/CDW boundary $V_c$ of the EHM with $\alpha_M = 2$ leads to 
$U(EHM) < U(EEHM) \approx U(PCM) \approx 0.61$ eV that we prefer on the basis of $\alpha_M \approx 1.5$. 
The Na and K salts are in the CDW phase with $V > V_c$ while $\chi(T)$ of Rb-TCNQ(II) 
is consistent with a BOW phase with $V$ close to $V_c$. The $\chi(T)$ fit of Na leads 
to $\delta = 0.2$ up to $T \approx 310$ K and the interpolation in Eq. \ref{eq3} 
for coexisting phases up to $T_d$. The K-TCNQ fit has larger $\delta = 0.4$ at low $T$. 
The knee region of Rb-TCNQ(II) is fit by Eq. \ref{eq5} with $2\xi = 60$, the soliton width used previously \cite{r30} 
for IR data.

\section{Model parameters }
The parameters in Table \ref{tb3} are for three models 
with BOW and CDW phases. They  are internally consistent, 
but considerably smaller than expected from optical data. Typical values \cite{r1,r2,r3,r4} are 
$t \approx 0.1 - 0.3$ eV and $U \approx$ 1 eV 
in Hubbard models or larger $U_e = U - V \approx 1$ eV in the EHM. Such parameters rationalize a CT transition around 1 
eV polarized along the stack and magnetic excitations at lower energy $4t^2/U_e \approx 0.1$ eV. 
We return in the Discussion to model parameters. Here we report direct evaluation of $U$, $V$ and $t$ 
for individual or adjacent $ \rm TCNQ^-$.  
The results are based on density functional theory (B3LYP) with the 6-311**G(p,d) 
basis in the Gaussian 03 package \cite{r47}. An eclipsed  $ \rm (TCNQ)_2^{-2}$ 
dimer at R = 3.2 or 3.4 $\AA$  is correctly found to have singlet gs, while 
smaller basis sets \cite{r48} yield a triplet gs. Smaller basis sets are adequate for model parameters, 
however, as discussed \cite{r49} for  $t$.\\

The disproportionation reaction $2A^- \rightarrow A^{2-} + A$ relates 
 $U$ to the gs energies $E_0(A^-), E_0(A)$ and $E_0(A^{2-})$. 
The optimized $\rm TCNQ^-$ structure leads to $U$(vertical) = 4.413 eV. 
Optimization of TCNQ and $\rm TCNQ^{2-}$ returns $U$(adiabatic) = 4.192 eV. 
The relaxation energy of 0.22 eV for electron transfer is in excellent 
agreement with 0.1 eV per $\rm TCNQ^-$ deduced \cite{r43} from Raman and IR spectra. 
The interaction $V$ depends on adjacent $\rm TCNQ^-$ and can be estimated several 
ways: (1) electrostatic repulsion between the atomic charges $q_i$ of the two ions; 
(2) repulsion between $q_i$ obtained in a dimer calculation; 
(3) energy difference $^3E_0 - 2E_0(A^-)$ between the triplet gs of the dimer, 
which precludes the formation of a $\pi -\pi$  bond, and two radical ions. 
The same values are obtained \cite{r48} to better than 5\%, and $V$’s in Table \ref{tb4} are based on the triplet. 
The listed $V(Na)$ and $V(K)$ are for eclipsed $\rm (TCNQ)_2^{2-}$ with R = 3.385  and 3.479$\AA$, 
respectively. The regular Rb-TCNQ(II) stack has R = 3.241 $\AA$ and a 2.0 $\AA$ displacement 
along the long axis shown in Fig. \ref{fig1}.\\ 

The ratio $U(ad)/V \approx 1.5$ is comparable to $\alpha_M = 1.6$ 
for  $\gamma = 0.2$ in Eq. \ref{eq1} or to $\alpha_M(8) = 17/12$ for an 8-site PCM. 
The 1D stack is close to the CDW boundary of the EEHM or PCM. The magnitude of $U$ 
is strongly reduced in the solid state by electronic polarization $P \approx 1$ eV 
per charge \cite{r50}. Since $P$ is approximately quadratic in charge, we have $P(A^{2-}) - 2P(A^-) \approx 2$ eV. 
Electronic polarization of adjacent $A^-$ reduces $V$ by a smaller amount.\\ 

The $t$’s in Table \ref{tb4} are for the 300 K structure of Rb and the $T > T_d$ structures of Na and K. 
We find $t(Na) > t(K) > t({\rm Rb})$ as expected but for calculated $t$’s that exceed the 
magnetic parameters in Table \ref{tb3} by a factor of 2.5 for Rb and 3.5 for Na or K. The reason for 
such large reduction is not understood. There are two dimerized stacks \cite{r5} in Na or K-TCNQ at 300 K. 
Table \ref{tb4} lists the calculated $t_1$, $t_2$ and the larger, smaller separation $\rm R_+$, $R_-$. 
We obtain $\delta(Na) = (t_2 - t_1)/(t_2 + t_1) = 0.22$ or 0.26, somewhat larger than 
$\delta = 0.20$ at low $T$ in Fig. \ref{fig3}. The corresponding $\delta(K)$ are 0.27 and 0.25, 
smaller than $\delta = 0.40$ in Fig. \ref{fig4}. But the K salt has substantially larger 
$(t_1 + t_2)/2 = 1.17t$ in the dimerized phase that leads to an equally good $\chi(T)$ for $\delta = 0.30$ 
when the mean value of the transfer integral is used. Overall, the calculated and fitted $\delta$ 
are reasonably consistent.\\ 
\begin{table}
\begin{center}
\caption {Calculated model parameters for adjacent $\rm TCNQ^-$}.
\begin{tabular}{c|c|c|c} \hline
Parameter & Na-TCNQ~~~~ & ~~~~ K-TCNQ ~~~~~& Rb-TCNQ(II)\\\hline
$V $(eV) & 2.713 & 2.671 & 2.594\\
$t $(eV) & 0.345 & 0.299 & 0.182\\
$t_1/t_2$ (eV) & 0.299 / 0.468 & 0.254 / 0.444 & -\\
$t_1/t_2$ (eV) & 0.266 / 0.451& 0.265 / 0.429 & -\\
$\alpha$ (eV/$\AA$) & 0.59 &  0.55 & 0.34 \\\hline

\end{tabular} 
\label{tb4}
\end{center}
\end{table}

The two-point derivative $dt/d{\rm R} = \alpha$ is an estimate for the e-ph coupling constant. 
The two stacks of Na or K-TCNQ at $T < T_d$ have almost the same $\alpha$, 
whose average value is reported in Table \ref{tb4}. 
The 300 and 100 K structures of Rb-TCNQ(II) return a smaller $\alpha \approx 0.34 eV/\AA$. 
The structural constraint of $2u < 0.04  \AA$ discussed above leads to $\delta(0) < \alpha u/t = 0.036$, 
an estimate that is independent of reduced $t$ in the crystal. Dimerization fluctuations 
of such small amplitude would be difficult to detect.\\ 

Direct evaluation of $t$ has been discussed before \cite{r49,r51}. Eclipsed $\rm TCNQ^-$ 
gives the largest $t$(R,0) that decreases with increasing separation R. 
Displacing the ion by L along the long axis leads to tilted stacks in Fig. \ref{fig1}. 
The nodes of the singly occupied orbital of $\rm TCNQ^-$ generate $t({\rm R,L}) = 0$ at  L=1.3$ \AA$ 
and to secondary maxima at other L \cite{r49,r52}. The first maximum at L = 2.1 $\AA$ 
 is close to the Rb-TCNQ(II) or TTF-TCNQ structures. 
A series of substituted perylenes illustrates wider variations of $t$ with displacements along 
both the long and short molecular axes \cite{r51}.\\

We consider next parameters derived from nonmagnetic data. Simple TCNQ salts have a broad CT absorption 
$\hbar \omega_{CT} \approx 1 eV$ polarized along the stack. The optical conductivity of K-TCNQ has a shoulder at higher energy that has 
been variously associated with dimerization \cite{r53}, with a band edge \cite{r54} or with a 
local excited state \cite{r45} of $\rm TCNQ^-$. Meneghetti \cite{r55} modeled   
K-TCNQ with special attention to totally symmetric mid-IR modes that are coupled in dimerized 
stacks to the CT absorption. Polarized spectra yield the coupling constants $g_n$. Meneghetti \cite{r55} 
used an EHM with $N = 4$ sites, periodic boundary conditions, and adjustable $t_1$, $t_2$ and $V_1$, $V_2$ at $T < T_d$. 
Comparison with experiment also entails lifetime or broadening parameters. 
Nearly quantitative fits are shown in Fig. 8 of ref. \cite{r55} for the optical 
conductivity at 300 K with coupled mid-IR modes and in Fig. 7 for polarized spectra at 27, 300 and 413 K. \\ 

The EHM parameters of ref. \cite{r55} for a regular K-TCNQ stack are $t = 0.19$ eV, $U = 1.20$ eV and $V = 0.02$ eV. 
Neglecting $V$ for a moment, we have a Hubbard model with $\chi(0) \approx  1.0 \times 10^{-4}$ emu/mole. 
Including $V = 0.02$ eV in a $N = 8$ calculation leads to $\chi(T)$ with a broad maximum at 1.6 x $10^{-4}$ emu/mole 
at $T_{max} \approx 800$ K, consistent with the magnitude of regular stacks in Fig. \ref{fig1}. 
The observed $\chi(T)$ slope for $T > T_d$ is much steeper, however, and finite $\chi(0)$ is not consistent with 
Rb-TCNQ(II). We note that $V = 0.02$ eV is a finite-size effect since $N = 4$ confines an e-h excitation to 
be close together. The CT absorption  shifts to lower energy with increasing N and optical spectra of 
longer regular stacks return different parameters. The 300 K parameters of ref. \cite{r55} are again $U = 1.20$ eV 
and alternating $t_1 = 0.10$, $t_2$ = 0.37 eV (or $\delta = 0.27/0.47 = 0.574$) and $V_1 = 0.28$, $V_2 = 0.31$ eV. 
The strong CT absorption at $N = 4$ hardly shifts to the red at $N = 8$ or 12. But $\delta = 0.574$ opens a large $E_m$. 
The calculated $\chi(T)$ for $N = 8$ with these parameters is very small ($<10^{-6}$ emu/mole) up to 500 K, 
completely incompatible with the magnetic data in Fig. \ref{fig1}. \\ 
 
Quantitative treatment of e-mv coupling in dimerized stacks such as K-TCNQ is based on 
linear response theory and force fields for molecular vibrations \cite{r56,r57}. The coupling constants $g_n$ 
 depend on just one electronic parameter, the zero-frequency optical conductivity. 
While the CT band is of central importance, its precise modeling is not. 
Dimerized stacks with broken $\rm C_i$ 
symmetry are required for coupling to mid-IR modes. Coupling to the {\it same } mid-IR modes 
in Rb-TCNQ(II) in regular stacks is strong evidence for a BOW phase with broken $\rm C_i$ symmetry. 
The same modes appear \cite{r58} with slightly higher intensity in powder spectra of Rb-TCNQ(I), which is 
dimerized \cite{r15} at 300 K. The $T$ dependence of the intensities $I_{\rm IR}(T)$ of coupled modes
 is characteristic of a BOW phase, and spin solitons with $2\xi = 60$ account \cite{r30} for $I_{\rm IR}(T)$.\\ 

The optical spectrum in the narrow BOW phase is dominated by $t$ due to competition 
between the larger $U$ and $V$ terms. The CT absorption peak is around $3t$ for a regular stack 
of rigid molecules \cite{r34} and shifts to higher energy by $U$(vert) - $U$(ad) = 0.22 eV. 
Dimerization also shifts $\hbar \omega_{CT}$  to higher energy. Preliminary modeling with all eigenstates of $N = 8$ or 
10 indicates that the $t$’s in Table \ref{tb3} have to be doubled for optical spectra and 
that $\delta \approx 0.3$ produces small blue shifts without a shoulder on the high-energy side. 
Larger $t$’s have been assumed all along for optical spectra.\\

\section {Discussion}

We have modeled the molar spin susceptibility $\chi(T)$ of alkali-TCNQ 
salts in Fig. \ref{fig1}  that were previously beyond quantitative treatment. We have not treated the phase 
transitions of Na or K-TCNQ, but relied on crystal data for $T_d$ and coexisting phases or 
diffuse scattering. We found consistent parameters in Table \ref{tb3} for 1D Hubbard models 
with point charges or second-neighbor $V$ that reduced the Madelung constant to $\alpha_M \approx 1.5$. 
The $\chi(T)$  fits in Fig. \ref{fig3},\ref{fig4},\ref{fig5} have $t$, $U$, $V$ parameters 
in Table \ref{tb3} that are about half as large as parameters from optical data.\\
    
It should perhaps be no surprise that quantitative analysis of magnetic and optical 
data within the same model leads to different parameters. Hubbard models make the zero-differential-overlap (ZDO) 
approximation of H\"{u}ckel theory for conjugated molecules or of tight-binding theory in solids. 
The PCM with $V_n = V/n$ is a special case of the Pariser-Parr-Pople (PPP) model \cite{r59,r60}. 
Salem \cite{r60} has summarized the merits and limitations of ZDO, which does not concern us here. 
But his discussion of $t$, the H\"{u}ckel $\beta$ parameter, bears directly on different magnetic 
and optical parameters. Systematic variations are illustrated by many conjugated 
hydrocarbons with $\rm sp^2$ hybridized C atoms. H\"{u}ckel theory provided a convenient approach 
to analyze variations prior to modern digital computers. Thermochemical data were 
successfully fit with a $\beta_{th}$ that is roughly half of $\beta_{op}$ inferred 
from optical spectra \cite{r60}. The correlated PPP model with $\beta_{op}$ is defined by 
the geometry of planar conjugated molecules and has considerable predictive power \cite{r57,r61}, 
including two-photon spectra and nonlinear optical properties. More recently, INDO 
(intermediate neglect of differential overlap) and its spectroscopic version INDO/S have 
different $\beta$ parameters \cite{r62}.\\ 

Instead of closely related hydrocarbons, Hubbard models are used to study electron-electron correlation 
solids in general. Quantitative application is rare and so are homologous series. Moreover, magnetic and 
optical or other properties are typically modeled separately and a single half-filled Hubbard band is 
rarely thought to be quantitative. Na, K and Rb-TCNQ(II) 
are closely related quasi-1D systems that nevertheless crystallize in different space groups. \\

At least qualitatively, differences between magnetic and optical parameters may be rationalized 
in terms of relaxed states in thermal equilibrium and electronic excitations that are fast compared to 
atomic or molecular motions. Equilibrium states that contribute to $\chi(T)$ are fully relaxed 
with respect to both molecular and lattice modes, and relaxed states have reduced excitation energies. 
Hubbard or other approaches to electronic excitations start with vertical $0-0$ excitations. 
 Electronic polarization reduces $U$ and $V$ significantly in the solid state, but this fast process is 
fully included in model parameters for optical spectra. The Holstein model \cite{r63} illustrates reduced $t$ due 
to linear coupling to a molecular vibration. Lattice phonons are considered in 1D for selected modes 
such as the Peierls mode but complete 3D relaxation is prohibitively difficult. Yet such 
relaxation is the most likely explanation for small parameters derived from $\chi(T)$ data. Quantitative 
modeling of the spin susceptibility clearly points to different magnetic and optical parameters for Na, K and Rb-TCNQ(II). 
The magnetism also indicates the Hubbard-type models for the Na and K salts at $T > T_d$ are in the CDW phase while the Rb(II) 
salts is in the BOW phase close to the CDW boundary.\\

\noindent {\bf Acknowledgements.}\\ 
ZGS thanks A. Girlando for access to unpublished IR spectra, 
A. Painelli for stimulating discussions about 1D models and their instabilities, and R. Pascal, Jr., 
for analysis of X-ray data. We gratefully acknowledge support for work by the 
National Science Foundation under the MRSEC program (DMR-0819860).
 
\end{document}